# Surfing Liquid Metal Droplet on the Same Metal Bath via Electrolyte Interface

Xi Zhao,[1,2*] Jianbo Tang,[3*] and Jing Liu,[1,2,3,a)]

[1]Key Laboratory of Cryogenics, Technical Institute of Physics and Chemistry, Chinese Academy of Sciences, Beijing 100190, China.

[2]School of Future Technology, University of Chinese Academy of Sciences, Beijing 100049, China.

[3]Department of Biomedical Engineering, School of Medicine, Tsinghua University, Beijing 100084, China.

* These two authors contributed equally to this research.

a) Corresponding author. Email: jliu@mail.ipc.ac.cn.

## Abstract:

We reported a phenomenon that when exerting an electric field gradient across a liquid metal/electrolyte interface, a droplet of the same liquid metal can persistently surf on the interface without coalescence. A thin layer of the intermediate solution, which separates the droplet from direct metallic contacting and provides the levitating force, is responsible for such surfing effect. The electric resistance of this solution film is measured and the film thickness is further theoretically calculated. The fact that the levitating state can be switched on and off via a controlled manner paves a way for reliable manipulation of liquid metal droplet.

## 1. Introduction

Intuition expects that when a droplet is gently placed on a surface of the same liquid, the droplet will coalesce instantly with the bulk.[1] Though, non-coalescence phenomenon is equally common but perhaps less noticed.[2,3] Isothermal non-coalescence of liquid drops over liquid surface can be traced back to 1879 reported by Lord Rayleigh,[4] and in 1881, Osborne Reynolds observed what he called floating drops.[5,6] The non-coalescent droplets have been formed by coating droplets with non-wetting hydrophobic powder,[7] thermo-capillary convection,[2,8] vertical oscillation of a liquid bath,[4] a high-speed moving surface,[9,10] and a sufficient hot surface.[11] In these levitation situations, the floating droplet avoids any contact with a solid element, thus the disappearance or reduction of friction with the substrate arouses remarkable dynamics of fluids which possesses great fundamental significance.[12]

Here we report an unconventional surfing effect of non-coalescent liquid metal droplet (LMD) inside a NaOH solution triggered by an electric field. Compared to traditional nonmetallic materials, gallium-based room temperature liquid metal (LM) owns totally different characters. It has much higher density and surface tension than that of water and silicone oils, which are commonly used to observe the non-coalescence phenomenon.[13] Distinct from previous experiments performed in air, the surfing phenomenon in this study takes place at the LM-electrolyte interface which puts forward a levitation method for on-demand manipulation of LM motors.[14] Also, it provides further understanding of the abundant interfacial phenomena of the LM.[15-17]

## 2. Material and Phenomena

The LM and the electrolyte used in this study are eGaIn (Eutectic Gallium and Indium, 75.5% gallium and 24.5% indium by weight percent) and NaOH solution ($0.25\ \mathrm{mol/L}$), respectively. During the experiment, a cathode is inserted in the lower LM bulk and an anode is placed in the upper electrolyte. An interfacial flow is initiated instantly when a voltage is applied between the electrodes. When depositing a LMD on the rapid-flowing interface, it levitates and surfs along with the flow. Fig. 1(a) presents the side view of surfing LMDs with various sizes captured with a high-speed camera (IDT, NR4.S3, 200 frame/s). The capillary length[11,18] for eGaIn in 0.25 mol/L NaOH solution is $\ell_c = 3.05\ \mathrm{mm}$, corresponding to a critical droplet volume of $120\ \mu\mathrm{l}$. LMDs either below or surpass this critical volume (from $6\ \mu\mathrm{L}$ to $3000\ \mu\mathrm{L}$) are both able to maintain long-term non-coalescent which means the surfing phenomenon is not volume-sensitive. The phenomenon, however, takes place only in the presence of an applied electric field gradient across the LM-electrolyte interface. Once the electric field is removed, the surfing droplet immediately merges with the LM bulk (Fig. 1(b)). The levitation provided by the deformed LM/electrolyte interface can be so strong that it can hold an impacting droplet (Fig. 1(c)). In addition, successive adding of LMDs can form a cluster of LMDs gathering and surfing together (Fig. 1(d)). It is found that the cluster will not coalesce with the underneath LM bath. Instead, they tend to merge with their neighbors to form bigger ones. The surfing LMDs and the underneath LM bulk look like two independent parts which do not disturb each other.

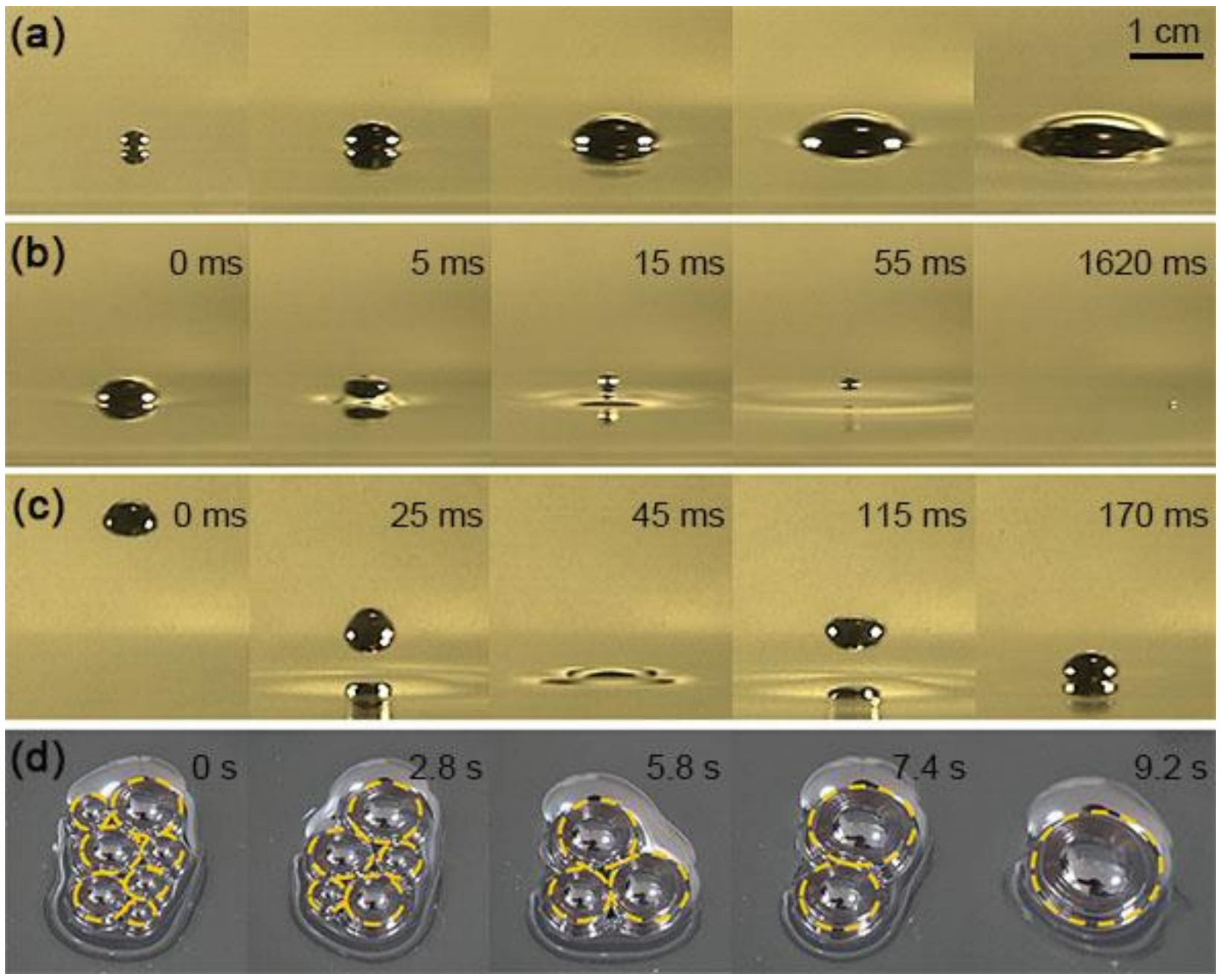

FIG. 1. (a) Surfing droplets with various sizes; (b) A droplet merges with the LM bulk immediately as the applied voltage is switched off; (c) An impacting non-coalescent LMD drops from $5\ \mathrm{cm}$ height; (d) A surfing LMD cluster coalesces into a large droplet.

## 3. Results and Discussion

It is obvious that the electric field plays a decisive role in our experiments. To further find out the influences of electric field on the surfing phenomenon, the voltage-dependence of the maximum volume of non-coalescent LMD $V_{\mathrm{M}}$ is examined. It can be seen from Fig. 2(a) that the LMD can only maintain non-coalescent when the applied voltage $U$ is higher than $2.8\ \mathrm{V}$. And $V_{\mathrm{M}}$ grows gradually as $U$ increases until $4.5\ \mathrm{V}$, after which $V_{\mathrm{M}}$ becomes almost identical. The existence of the voltage threshold for the surfing effect is in agree with of our experimental observation that the LM/electrolyte interface starts to flow only when $U$ reaches $2.8\ \mathrm{V}$ or higher, which is a mandatory requirement for the surfing effect according to previous studies.[19-21] The flow is induced by the electric gradient across the two immiscible fluids of distinct electric conductivity (LM- $3.40 \times 10^6\ \mathrm{S/m}$, 0.25 mol/L NaOH- $5.3\ S/m$) according to electrohydrodynamic principles.[22,23] The mean flow velocity $v$ corresponding to different applied voltage ranging from 0 V to 8 V is measured and presented in Fig. 2(b). It can be found that $v$ shows a quite similar voltage-dependence to $V_{\mathrm{M}}$. The velocity measured during the tested voltage range is on the order of several centimeters per second.

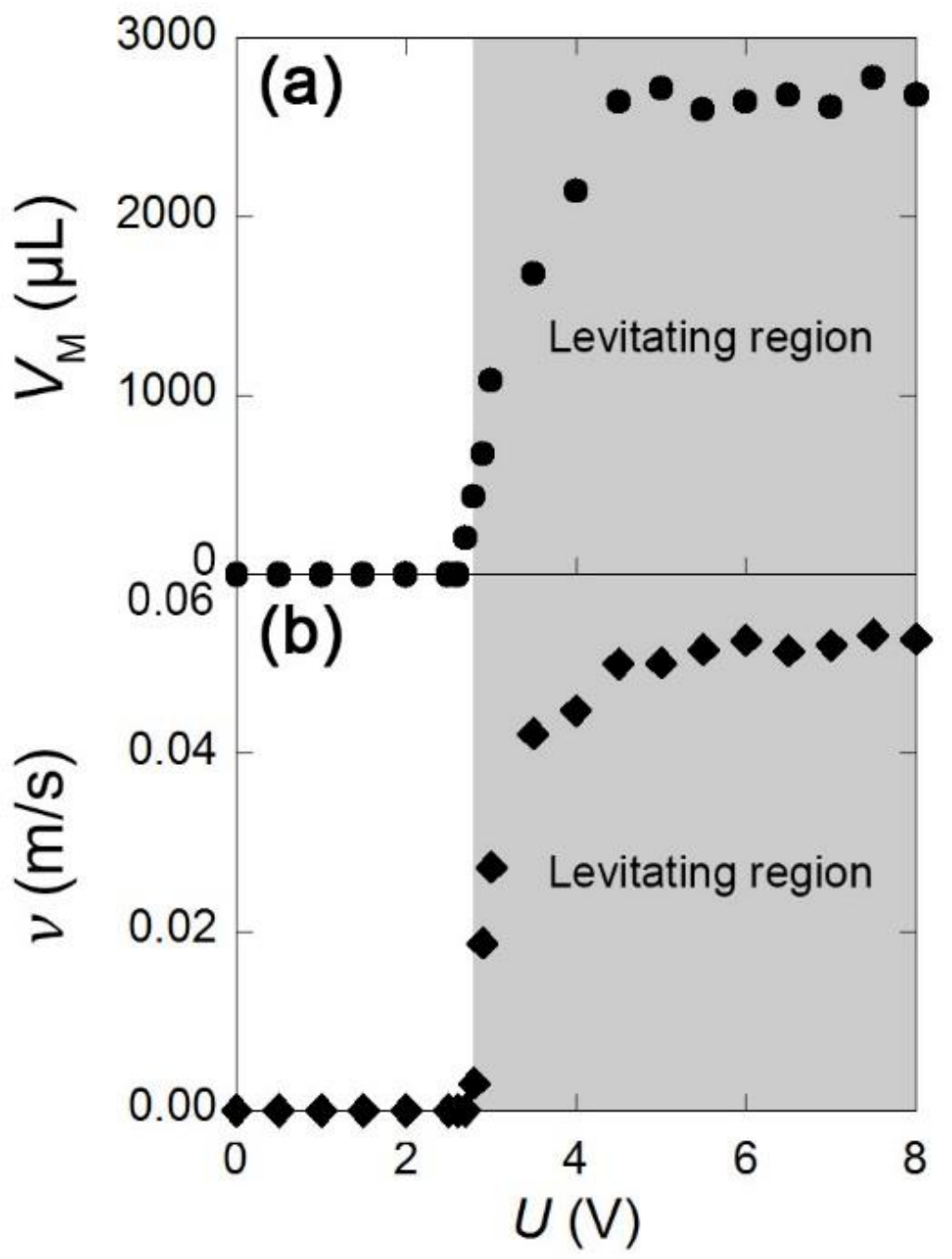

FIG. 2. (a) The maximum volume of non-coalescent LMD $V_M$ under different applied voltage $U$; (b) The flow velocity $v$ of the LM-electrolyte interface under different $U$. The velocity is measured by tracing tiny particles dispersed on the interface with particle image velocimetry technique.

In contrast to conventional gas-film levitating cases,[9,24] the surfing mechanism of current study is ascribed to a liquid film (NaOH solution) between the LMD and the LM bulk. This can be verified through measuring the electrical resistance $R$ between the LMD and the LM bulk since inserting a dielectric film between two excellent conductors will cause a significant increase of $R$. The evolution of $R$ before and after cutting off the applied voltage $U$ is measured and shown in Fig. 3(a). During the surfing state (Fig. 3(a)-i), when the LMD is levitated with an applied voltage of $U$, typical resistance values of several hundred ohms are measured. After terminating $U$, the absence of an electric field makes the flow slow down. However, the LMD will not coalesce with the bath immediately since the flow is not terminated immediately due to inertia. Because the resistance of a solution correlates closely to flow velocity[25,26], the flow deceleration brings a resistance decrease of about one order of magnitude to this pre-coalescence state (Fig. 3(a)-ii). Right after the pre-coalescence state, $R$ further drops below one ohm, indicating a direct metallic contact (coalescence state) of the LMD and bath underneath (Fig. 3(a)-iii). Droplets usually perform several similar partial-coalescence steps before being completely swallowed.[27,28] The repeated developing pattern of $R$ implies the LMD in our circumstance also follows this rule and experiences a second coalescence. The difference of $R$ between the pre-coalescence state and the coalescence state shows unambiguously that the existence of a liquid film do increase the resistance between the two LM bodies. And the resistance of the system should in return give information about the film. To eliminate the influence of the motion of the solution on its resistance, the resistance at the quasi-stable pre-coalescence state $R_{\mathrm{ii}}$ is used for later comparisons. It can be found in Fig. 3(b) that an increase in droplet size will decrease the resistance of the solution film as the droplet volume ranges from $20\,\mathrm{uL}$ to $1000\,\mathrm{uL}$. The resistance $R_{\mathrm{ii}}$ for small LMDs $(\mathrm{Bo} < 0.6)$ which can be treated as a sphere is magnified in the inset of Fig. 3(b) for the coming discussion.

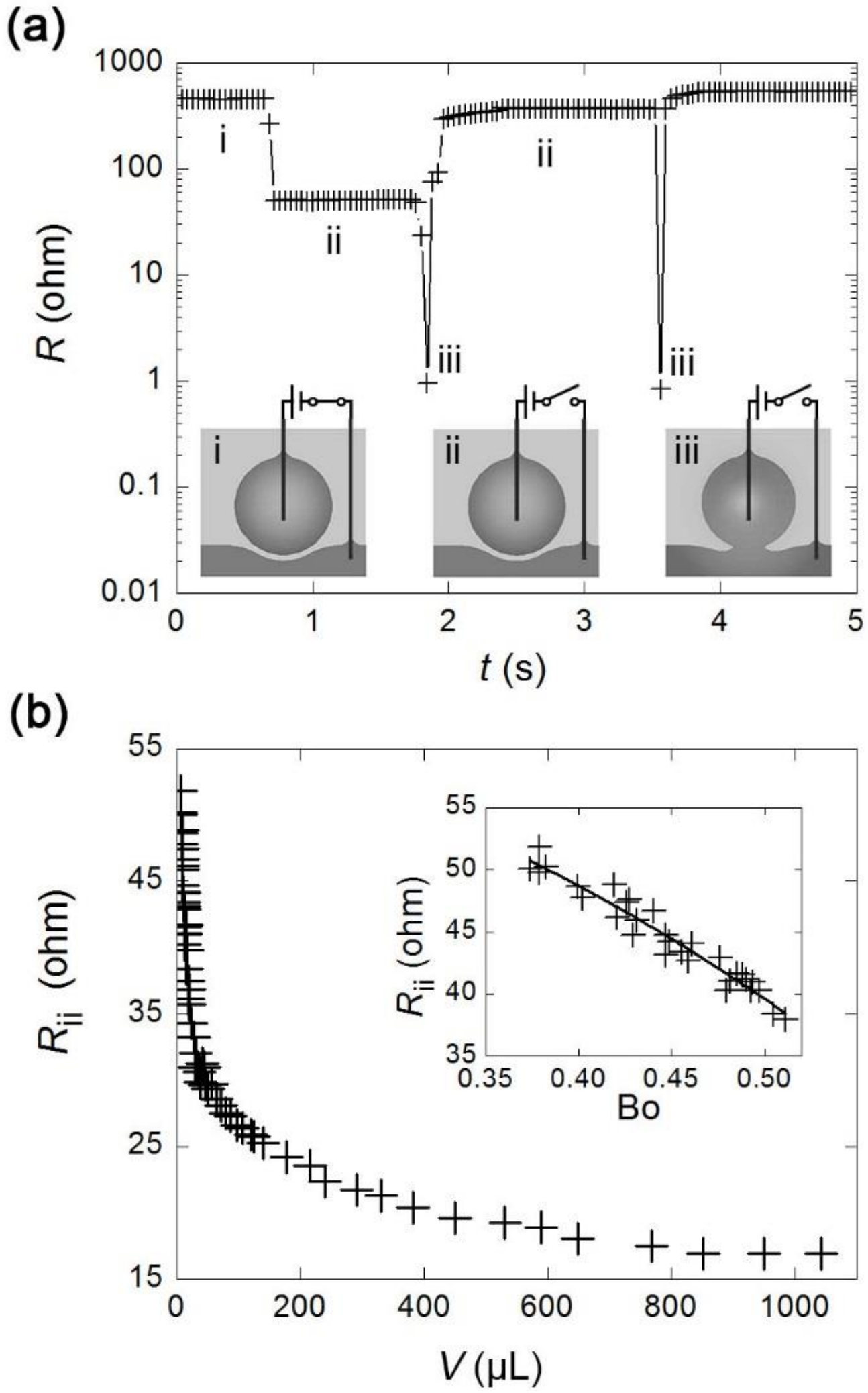

FIG. 3. (a) Resistance evolution of the solution film before and after cutting off the applied voltage. Insets: i-surfing state, ii-pre-coalescence state, iii-coalescence state. The difference of $R_{\mathrm{ii}}$ between the first and the latter pre-coalescence states is attributed to the increase of film thickness when the droplet is partially coalesced; (b) $R_{\mathrm{ii}}$ decreases as the droplet volume increases. Inset: $R_{\mathrm{ii}}$ of LMDs for $\mathrm{Bo} < 0.6$.

The thickness of the bearing film $e_0$ is a key parameter of non-coalescence phenomenon. When transparent fluids and gas bearing films are engaged, $e_0$ can be experimentally measured through optical interference.[4,10] However, when it comes to the non-transparent LM, direct measurement of $e_0$ is inaccessible. Fortunately, the unique merit of our case that the bearing film, which is a solution by nature, enables us to estimate $e_0$ based on $R_{\mathrm{ii}}$ by taking into consideration of the geometrical configuration of the problem. The electric field lines for an equipotential spherical LMD separated by a thin solution film (thickness $e_0$) from the LM bath (equipotential plane) are sketched in Fig. 4(a). Since charges move along the electric field lines, the resistance of the system $R_{\mathrm{ii}}$ can be divided into two parts which are in parallel connection according to its electric field distribution. The first part is the bearing film which is treated as a spherical dome with a uniform thickness $e_0$ and the aperture of the spherical dome is $2\theta_{\mathrm{S}}$, where $\theta_{\mathrm{S}}$ is the separation angle measured from the bottom of the drop to the separation point.[11,29] The electric field lines within this zone have a straight radial distribution which results in a film resistance of:

$$R_1 = \frac{k e_0}{4\theta_s r^2} \tag{1}$$

where $k$ is the electrical resistivity of the solution and $r$ is the drop radius. The other part is the rest of the solution that surrounds the LMD in which the electric field lines are piecewise straightened for simplicity. Therefore, the equivalent resistance $R_2$ of this part can be approximated by integrating the infinitesimal shells from $\theta_{\mathrm{S}}$ to $\pi/2 + \theta_{\mathrm{S}}$ (See Supplemental Materials[30] for derivation details). In doing so, assumptions are made that the resistance contribution of the upper part of the solution (from $\pi/2 + \theta_{\mathrm{S}}$ to $\pi$) is marginal. A comparison between $R_1$ and $R_2$ shows that the former is always $14$ orders smaller than the later for LMD of varied sizes. Consequently, the parallel resistance of the two $R_{\mathrm{ii}} = 1/(1/R_1 + 1/R_2) \approx R_1$, which results in

$$e_0 = \frac{4\theta_s r^2 R_{\mathrm{ii}}}{k} \tag{2}$$

According to Eq. (2), the film thickness $e_0$ can be determined since the geometrical configurations of the LMD as well as $R_{\mathrm{ii}}$ are both accessible. Fig. 4(b) presents the measured $\theta_{\mathrm{S}}$ as a function of $r$ for droplets whose $\mathrm{Bo} < 0.6$ and $\theta_{\mathrm{S}}$ is found to grow along with the increase of droplet size. Incorporating the already-known $R_{\mathrm{ii}}$, $\theta_{\mathrm{S}}$ and taking $k = 0.19\,\Omega \cdot \mathrm{m}$ for $0.25\,\mathrm{mol/L}$ NaOH solution, $e_0$ can be calculated based on Eq. (2). As shown in Fig. 4(c), the thickness of the bearing film ranges from several hundreds of micrometers to millimeter scale.

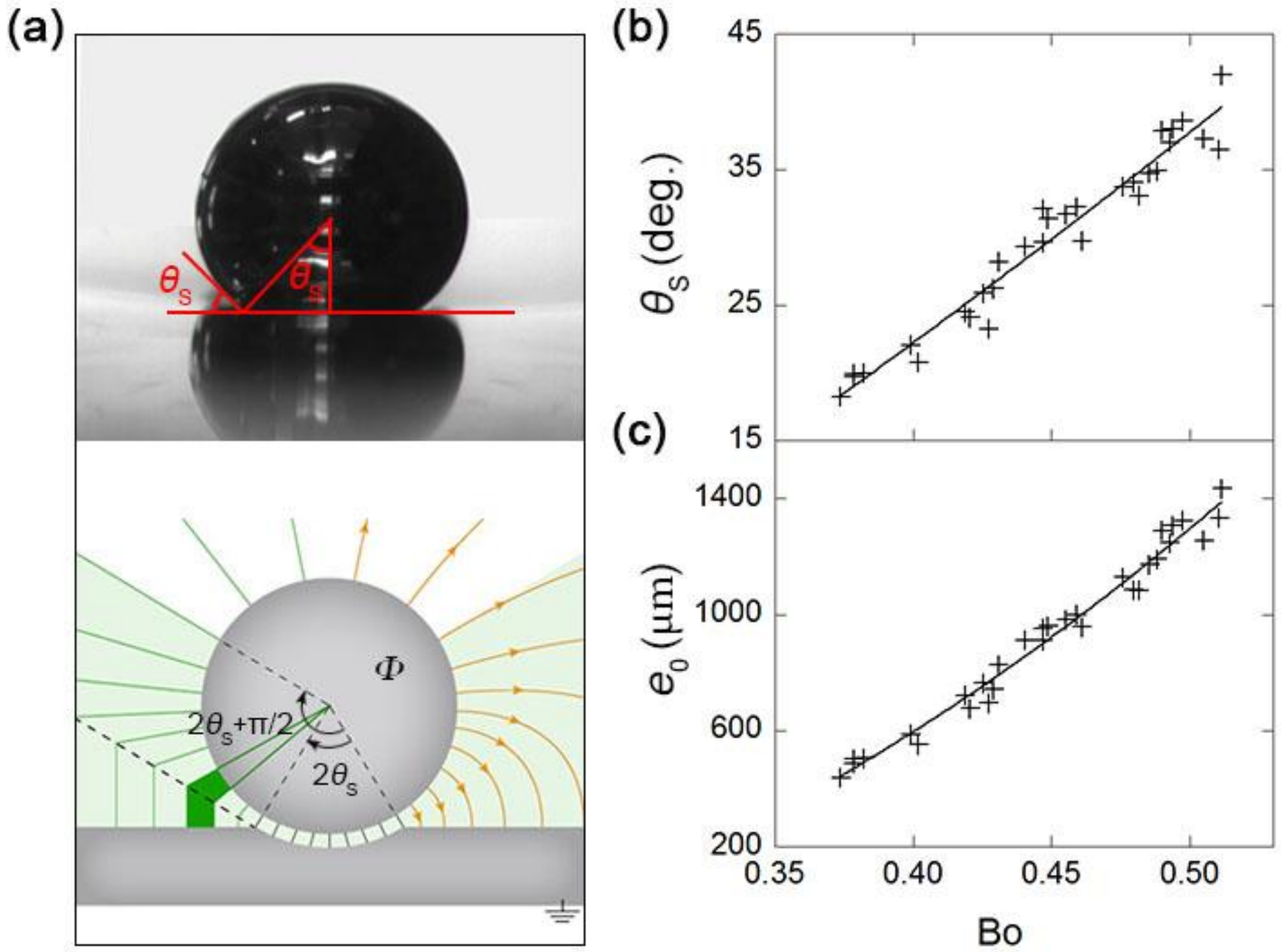

FIG. 4. (a) Typical side view of a surfing LMD (Top) and the simplified model for calculating $R_{\mathrm{ii}}$ (Bottom). The orange lines on the right depict the direction of the electric field while the piecewise green lines on the left represent their approximations. The dark-green-filled area represents the cross section of an integral infinitesimal (cylinder shell) and the light-green-filled area represents the entire integral domain, respectively. (b) Separation angle $\theta_{\mathrm{S}}$ versus Bond number Bo; (c) Film thickness $e_0$ deduced from the resistance model.

The thickness of the film $e_0$ can also be estimated through a hydrodynamic approach by balancing the gravity $F_{\mathrm{G}} = 4/3\,\pi r^3 \rho_{\mathrm{LM}} g$ ($\rho_{LM} = 6280\ \mathrm{kg/m^3}$ is the density of eGaIn) with the buoyancy term $F_B = \pi(\rho_{\mathrm{LM}} - \rho_{\mathrm{E}}) g r^3 C_0$ ($\rho_{\mathrm{E}} = 1009.5\ \mathrm{kg/m^3}$ is the density of the solution, $C_0$ is a constant related to the system)[29,31] and the levitating force $F_{\mathrm{L}}$ provided by the film. According to the Reynolds bearing theory,[4,32] $F_{\mathrm{L}}$ is on the order of $24\pi\mu_{\mathrm{E}} v r^3/{e_0}^2$ ($\mu_{\mathrm{E}} = 1.08 \times 10^{-3}\ \mathrm{Pa \cdot s}$ is the dynamic viscosity of the solution). For small Bo, the film thickness generated by this method is of the same order as that of the electric resistance model (For instance, $e_0 = 138\ \mu\mathrm{m}$ when $\mathrm{Bo} = 0.37$) which confirms our theory. And both methods suggest that the liquid bearing film in the current case is much thicker than the gas-bearing situations.[4,9,32] But as Bo increases, the $e_0$ calculated from the resistance model deviates from the hydrodynamic method by showing a larger slope. There are several factors may contribute to this deviation. Both the resistance of the film and the bearing force are sensitive to the geometry of the deformed interface, which requires a proper determination of $\theta_{\mathrm{s}}$ and $r$. In fact, the real separation point (beyond which equal film thickness will not hold) is usually lower than expected, which will be whelmed by the concave meniscus of the deformed interface. And this results in lager values during the measurements of $\theta_s$ and $r$, which in turn produces a larger $e_0$.[4] For larger Bo, the LMDs have greater tendency to establish a non-spherical shape, making it even more difficult to determine the parameters. That is why $e_0$ shows better consistency for small Bo.

## 4. Conclusion

In summary, we have disclosed an abundant dynamics of the surfing effect of non-coalescent LMD on the LM/solution interface triggered by an electric field. The bearing electrolyte film filling between the LMD and its bath represents a major distinction to previous studies in which a gas film is usually involved. A theoretical model which links up the resistance and film thickness is established to clarify the surfing effect and the characteristic parameters of the system. The film thickness calculated from the resistance model and the hydrodynamic approach for small drops is found to be on the same order. The surfing effect revealed here provides a configuration for manipulation and study of LMs droplet-on-a-soft-substrate.

## Acknowledgments:

This work is partially supported by the Beijing Municipal Science and Technology Funding (Grant No. Z151100003715002) and Dean's Research Funding and the Frontier Project of the Chinese Academy of Sciences.

## Reference

[1]T. Gilet, N. Vandewalle, and S. Dorbolo, Phys. Rev. E **76**, 035302 (2007)..

[2]G. P. Neitzel and P. Dell'Aversana, Annu. Rev. Fluid Mech. **34**, 267 (2002).

[3]E. Honey and H. Kavehpour, Phys. Rev. E **73**, 027301 (2006).

[4]Y. Couder, E. Fort, C. H. Gautier, and A. Boudaoud, Phys. Rev. Lett. **94**, 177801 (2005).

[5]R. Savino, D. Paterna, and M. Lappa, J. Fluid Mech. **479**, 307 (2003).

[6]P. Dell'Aversana and G. P. Neitzel, Physics Today **51**, 38 (2008).

[7]P. Aussillous and D. Quéré, Nature **411**, 924 (2001).

[8]P. Dell'Aversana and G. P. Neitzel, Exp. Fluids **36**, 299 (2004).

[9]K. R. Sreenivas, P. K. De, and J. H. Arakeri, J. Fluid Mech. **380**, 297 (1999).

[10]YoshiyukiTagawa, TuanTran, and ChaoSun, J. Fluid Mech. **733**, 302 (2013).

[11]Y. Ding and J. Liu, Appl. Phys. Lett. **109**, 1153 (2016).

[12]A. Duchesne, C. Savaro, L. Lebon, C. Pirat, and L. Limat, arxiv **1301** (2013).

[13]D. Zrnic and D. S. Swatik, J. Less Common Metals **18**, 67 (1969).

[14]J. Zhang, Y. Yao, L. Sheng, and J. Liu, Adv. Mater. **27**, 2648 (2015).

[15]L. Hu, L. Wang, Y. Ding, S. Zhan, and J. Liu, Adv. Mater. **28**, 9210 (2016).

[16]B. Yuan, L. Wang, X. Yang, Y. Ding, S. Tan, L. Yi, Z. He, and J. Liu, Advanced Science **3**, 1600212 (2016).
[17]J. Tang, X. Zhao, J. Li, Y. Zhou, and J. Liu, Advanced Science **4**, 1700024 (2017).
[18]W. Bouwhuis, K. G. Winkels, I. R. Peters, P. Brunet, M. D. Van, and J. H. Snoeijer, Phys. Rev. E **88**, 023017 (2013).
[19]Schwabe, D. Hintz, and Peter, J. Non-Equilib. Thermodyn. **25**, 215 (2001).
[20]H. Linke, B. J. Alemán, L. D. Melling, M. J. Taormina, M. J. Francis, C. C. Dow-Hygelund, V. Narayanan, R. P. Taylor, and A. Stout, Phys. Rev. Lett. **96**, 154502 (2006).
[21]E. Yakhshi-Tafti, H. J. Cho, and R. Kumar, J. Colloid Interface Sci. **350**, 373 (2010).
[22]C. H. Chen, H. Lin, S. K. Lele, and J. G. Santiago, J. Fluid Mech. **524**, 263 (2005).
[23]A. I. Zhakin, Physics-Uspekhi **56**, 141 (2013).
[24]R. Shabani, R. Kumar, and H. J. Cho, Appl. Phys. Lett. **102**, 042105 (2013).
[25]W. J. Hamer and H. J. Dewane, in *Electrolytic conductance and the conductances of the halogen acids in water* (National Bureau of Standards, 1970).
[26]V. G. Levich and S. Technica, in *Physicochemical hydrodynamics*. (Prentice-hall, 1962).
[27]B. Yuan, Z. He, W. Fang, X. Bao, and J. Liu, Science Bulletin **60**, 648 (2015).
[28]S. Shim and H. A. Stone, Phys. Rev. Fluids **2**, 044001 (2017).
[29]A. V. Rapacchietta and A. W. Neumann, J. Colloid Interface Sci. **59**, 541 (1977).
[30]See supplemental material for derivation details of the resistance model.
[31]K. J. Baumeister and R. C. Hendricks, Adv. Cryog. Eng. **16**, 455 (1971).
[32]C. Pirat, L. Lebon, A. Fruleux, J. S. Roche, and L. Limat, Phys. Rev. Lett. **105**, 63 (2010).

# Supplemental material

Fig. S1 shows the theoretical model that we developed to calculate the resistance between the liquid metal droplet and the liquid metal bath $R$ by using the geometrical configurations of the problem. Since ions inside the solution move along the direction of the electric field lines between the two liquid metal bodies, $R$ can be calculated based on the electric field distribution. Also because the electric field lines should be perpendicular to the both of the outlines of the two liquid metal bodies which are equipotential, $R$ can be divided into two parts of solution regions connected in parallel. And for the sake of simplicity, the electric field lines are piecewise straightened (Fig. 4(a)). For $\alpha \in (-\theta_S,\ \theta_S)$, the region is simply treated as a film with a spherical dome and a uniform thickness of $e_0$, then the resistance of this region $R_1$ would be:

$$R_1 = \frac{ke_0}{4\theta_S r^2} \tag{1}$$

For $\alpha \in (\theta_S,\ \pi/2+\theta_S)$, The other region can be divided into thin solution shells and its resistance $R_2$ can be obtained by integrating the resistance of each shell (parallel resistors). Along the electric field direction, the length of each shell (the integral infinitesimal) $L_{(\alpha)}$, AB+BD, can be deduced based on the geometrical configurations:

$$L_{(\alpha)} = \mathrm{AB} + \mathrm{BD} = \frac{Y_0 - r\cos\alpha}{\cos\alpha} + \left(1 - \frac{1}{\cos\alpha}\right) Y_0 \tan\theta_S \boldsymbol{\varphi}(\alpha) \tag{2}$$

where $Y_0 = (r + e_0)\cos\theta_S$, $\boldsymbol{\varphi}(\alpha) = (\tan\alpha - \tan\theta_S)/(1 + \tan\theta_S \tan\alpha)$.

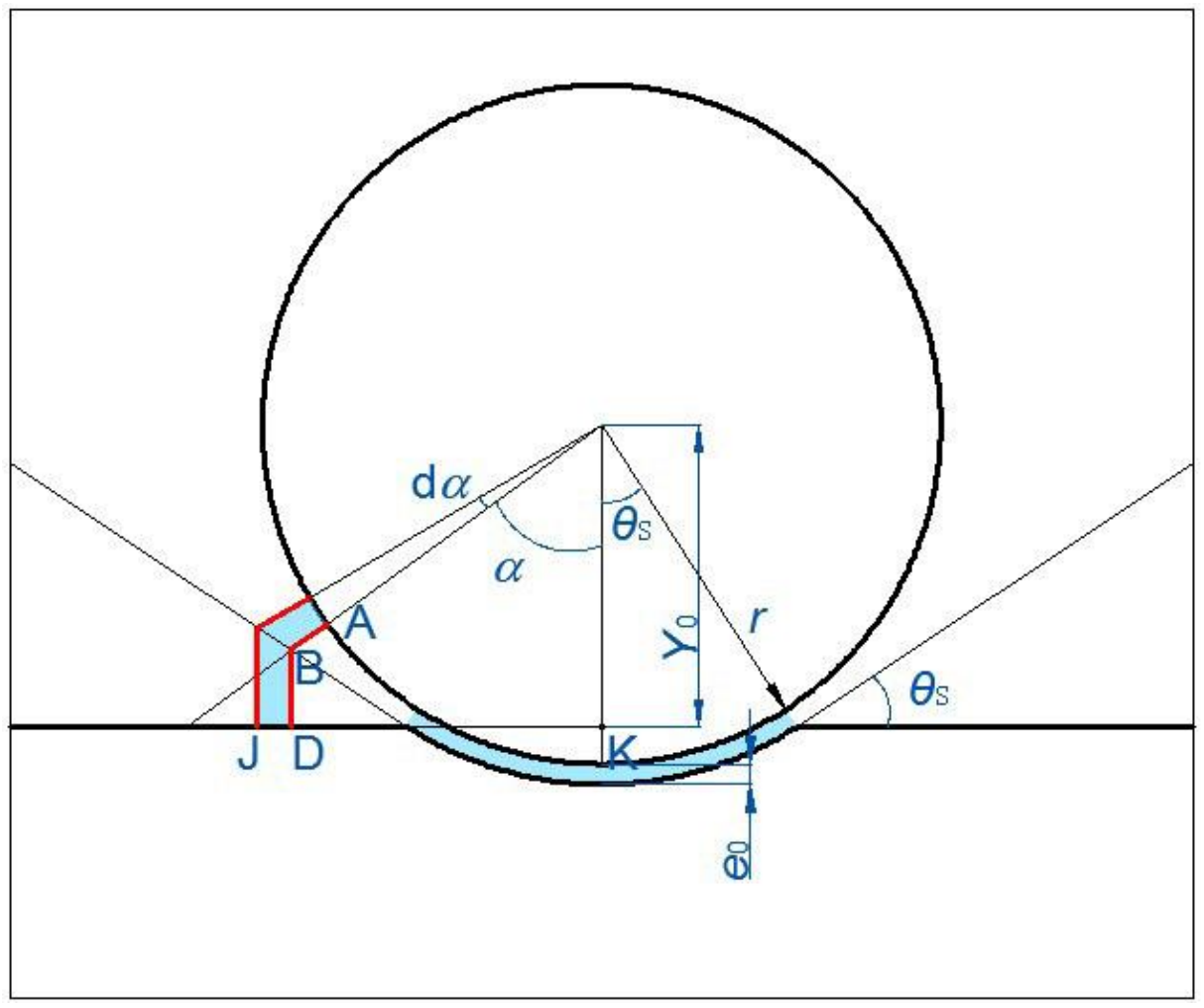

**Figure S1.** The theoretical model to calculate the electric resistance of the NaOH solution film.

The sectional area of each shell, $A_{(\alpha)}$ equals to:

$$A_{(\alpha)} = \pi(\mathrm{JK}^2 - \mathrm{DK}^2) = \pi {Y_0}^2[(\boldsymbol{\varphi}(\alpha + d\alpha) + \tan\theta_S)^2 - (\boldsymbol{\varphi}(\alpha) + \tan\theta_S)^2] \tag{3}$$

The resistance of each shell, $R_{(\alpha)}$, therefore, should be:

$$R_{(\alpha)} = \frac{kL_{(\alpha)}}{A_{(\alpha)}} = \frac{k}{\pi Y_0} \frac{\frac{1}{\cos\alpha} - \frac{r}{Y_0} + \left(1 - \frac{1}{\cos\alpha}\right) \tan\theta_S \boldsymbol{\varphi}(\alpha)}{(\boldsymbol{\varphi}(\alpha + d\alpha) + \tan\theta_S)^2 - (\boldsymbol{\varphi}(\alpha) + \tan\theta_S)^2} \tag{4}$$

Thus $R_2$ can be obtained by integrating $1/R_{(\alpha)}$ using an integral interval of $(\theta_S, \pi/2 + \theta_S)$. In doing so, we neglect the contribution of the solution shells to $R_2$ from $(\pi/2 + \theta_S)$ to $\pi$ given such contribution decays as $\alpha$ increases, which produces:

$$\begin{aligned} \frac{1}{R_2} &= \int_{\theta_S}^{\frac{\pi}{2}+\theta_S} \frac{1}{R_{(\alpha)}} d\alpha = \frac{\pi Y_0}{k} \int_{\theta_S}^{\frac{\pi}{2}+\theta_S} \frac{(\boldsymbol{\varphi}(\alpha + \mathrm{d}\alpha) + \tan\theta_S)^2 - (\boldsymbol{\varphi}(\alpha) + \tan\theta_S)^2}{\frac{1}{\cos\alpha} - \frac{r}{Y_0} + \left(1 - \frac{1}{\cos\alpha}\right) \tan\theta_S \boldsymbol{\varphi}(\alpha)} d\alpha \\ &= \frac{\pi Y_0}{k} f_0 \end{aligned} \tag{5}$$

where $f_0$ is a constant which equals to:

$$f_0 = \int_{\theta_S}^{\frac{\pi}{2}+\theta_S} \frac{(\boldsymbol{\varphi}(\alpha + \mathrm{d}\alpha) + \tan\theta_S)^2 - (\boldsymbol{\varphi}(\alpha) + \tan\theta_S)^2}{\frac{1}{\cos\alpha} - \frac{r}{Y_0} + \left(1 - \frac{1}{\cos\alpha}\right) \tan\theta_S \boldsymbol{\varphi}(\alpha)} \mathrm{d}\alpha \tag{6}$$

which can be further simplified as:

$$f_0 = \int_{\theta_S}^{\frac{\pi}{2}+\theta_S} \frac{(1 + \tan^2\theta_S)^2 \cos^2\theta_S \left[\left(\frac{\sin(\alpha + \mathrm{d}\alpha)}{\cos(\alpha + \mathrm{d}\alpha - \theta_S)}\right)^2 - \left(\frac{\sin\alpha}{\cos(\alpha - \theta_S)}\right)^2\right]}{1 - \frac{r}{(r + e_0)\cos\theta_S} + \cos\theta_S (1 + \tan^2\theta_S) \frac{1 - \cos\alpha}{\cos(\alpha - \theta_S)}} \mathrm{d}\alpha \tag{7}$$

Finally, the total resistance between the surfing liquid metal droplet and the liquid metal bath, $R_{ii}$ can be obtained from Eq. (1) and Eq. (5):

$$R_{ii} = \frac{1}{\frac{1}{R_1} + \frac{1}{R_2}} = \frac{1}{\frac{4\theta_S r^2}{k e_0} + \frac{\pi f_0 (r + e_0)\cos\theta_S}{k}} \tag{8}$$

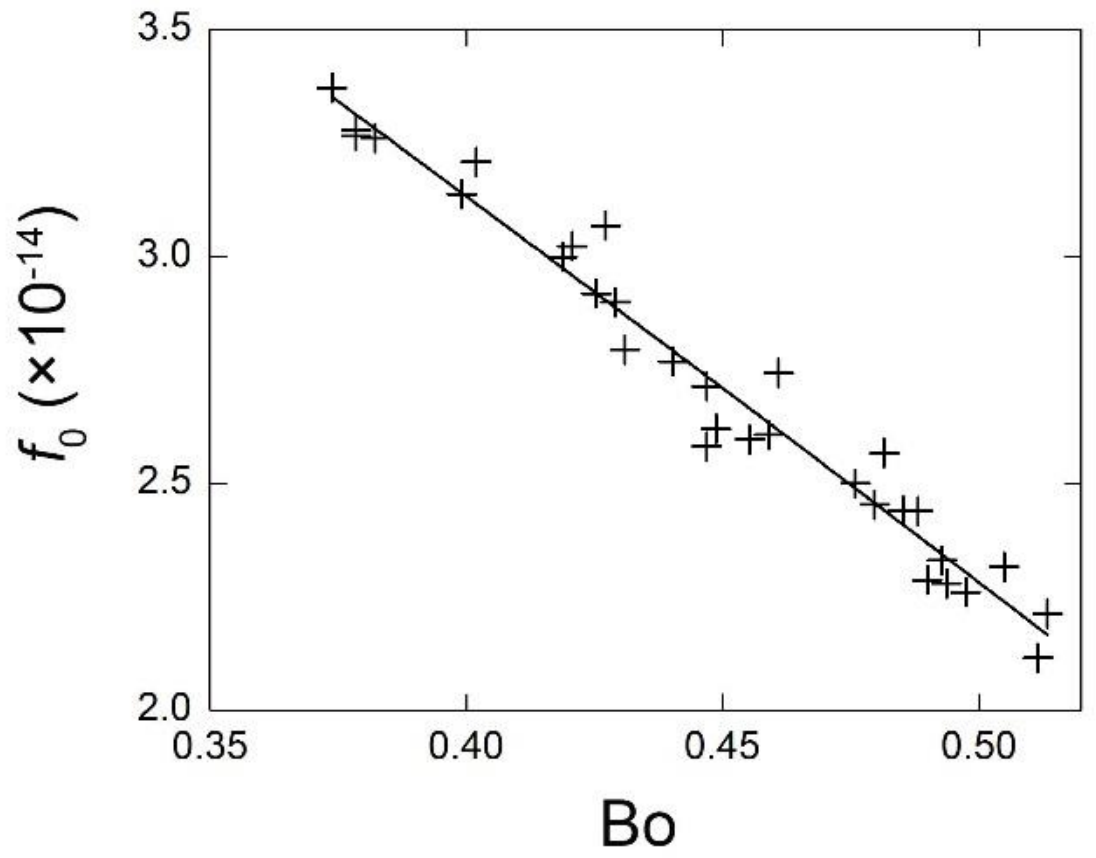

**Figure S2.** Calculated $f_0$ as a function of Bo.

A comparison between $R_1$ and $R_2$ can be deduced from Eq. (1) and Eq. (5):

$$\frac{R_1}{R_2} = \frac{\pi f_0 e_0 Y_0}{4\theta_\mathrm{S} r^2} = \frac{\pi \cos\theta_\mathrm{S}}{4\theta_\mathrm{S}} \cdot f_0 \left[\frac{e_0}{r} + \left(\frac{e_0}{r}\right)^2\right] \sim \frac{\pi f_0 \cos\theta_\mathrm{S}}{2\theta_\mathrm{S}} \cdot \frac{e_0}{r} \tag{9}$$

It can be seen from Figure S2 that the coefficient $f_0$ is very small (on the order of $10^{-14}$). Also given that $e_0 \ll r$, conclusion can be made that $R_1 \ll R_2$, which means $R_\mathrm{ii}$ is dominated by $R_1$:

$$R_\mathrm{ii} \approx R_1 = \frac{k e_0}{4\theta_\mathrm{S} r^2} \tag{10}$$